\documentclass[10pt]{iopart}
\usepackage{graphicx}
\usepackage{subfigure}
\expandafter\let\csname equation*\endcsname\relax
\expandafter\let\csname endequation*\endcsname\relax
\usepackage[square,sort&compress]{natbib}
\usepackage{array}
\newcommand{\ds}{\displaystyle}
\newcommand{\scs}{\scriptscriptstyle}

\def\<{\langle}
\def\>{\rangle}
\begin{document}
\title[Macroscopic transport equations from microscopic exclusion processes.]
{Nonlinear macroscopic transport equations in many-body systems from microscopic exclusion processes.} 
\author{Marta Galanti$^{1,2}$,
        Duccio Fanelli$^{1}$,
        Francesco Piazza$^2$}
\address{$^1$  Dipartimento di Fisica and INFN, Universit\`a di Firenze, I-50019, Sesto F.no (FI), Italy \\
         $^2$  Centre de Biophysique Mol\'eculaire, CNRS-UPR 4301, Universit\'e d'Orl\'eans, 45071 Orl\'eans cedex, France \\
         }
\ead{Francesco.Piazza@cnrs-orleans.fr}
%
%
%
\begin{abstract} 
Describing particle transport at the macroscopic or mesoscopic 
level in non-ideal environments poses fundamental theoretical 
challenges in domains ranging from inter and intra-cellular transport in biology to 
diffusion in porous media. Yet, often the nature of the constraints coming from many-body interactions 
or reflecting a complex and confining environment are better understood and modeled at the microscopic level.

In this paper we investigate the subtle link between microscopic exclusion processes and the 
mean-field equations that ensue from them in the continuum limit. We derive a generalized nonlinear 
advection diffusion equation suitable for describing transport in a inhomogeneous medium in the presence 
of an external field. Furthermore, taking inspiration from a recently introduced exclusion process
involving agents with non-zero size, we introduce a modified diffusion equation appropriate for describing 
transport in a non-ideal fluid of $d$-dimensional hard spheres. 

We consider applications of our equations to the problem of diffusion to an absorbing sphere 
in a non-ideal self-crowded fluid and to the problem of gravitational sedimentation. We show that 
our formalism allows one to recover known results. Moreover, 
we introduce the notions of {\em point-like} and {\em extended} crowding, which specify distinct 
routes for obtaining macroscopic transport equations from microscopic exclusion processes.
\end{abstract} 


\vspace{2pc}

\submitto{\JPA}

\maketitle

\section{\label{s:0}Introduction}

Diffusive transport is central in many areas of physics, chemistry, biology and 
soft matter~\cite{Crank:uq,Dhont1996,McGuffee:2010tg,Caspi:2002zr}.
However, while the mathematics of diffusive processes in dilute and simple media
is fairly well developed and understood~\cite{Crank:uq}, many interesting and relevant 
diffusive processes take place in  strongly non-ideal conditions. These include a wealth of different 
highly dense media, from non-ideal plasmas~\cite{kremp:216} to biological membranes~\cite{Saffman01081975}, 
media with complex topological structures, including porous media~\cite{Talmon:2010bu,Mitra:1992pt,Whitaker:1967ta} 
and living cells~\cite{Dix:2008lp,Phair:2000wa} and strongly confining environments~\cite{Condamin:2007ez,Zhou:2004km,Minton2000a,Zimmerman1993,Kim:2009fk,McGuffee:2010tg,Ando:2010kl}.

Crowding and confinement effects on diffusion-influenced phenomena still pose fundamental
yet unanswered questions. Concerning molecular mobility, for example, several computational and experimental indications
exist of anomalous diffusion in the cell cytoplasm as a result of amount and type of crowding~\cite{Balbo:2013ov,Weiss:2004vn},
suggesting that living cells behave much like fractal or otherwise disordered 
systems~\cite{ben-Avraham:2000kx,Bouchaud:1990ys}.
However, strong evidences also exist in favour of normal (Brownian) diffusion,  crowding 
and confinement resulting in this scenario in (often nontrivial) modifications of the diffusion 
coefficient~\cite{Dix:2008lp,Novak2009758,Konopka01092006,Nicholson01121981}. Another related issue is that of 
diffusion-limited reactions~\cite{Smoluchowski:1916fk}, which are ubiquitous in many domains in 
biology and chemistry, touching upon problems such as association,  
folding and stability of proteins~\cite{Zhou:2004oq,Cheung2005} and bimolecular reactions in 
solution~\cite{Dzubiella2005,Piazza2005,Fanelli:2010or,Schmit2009,Tachiya2007}, including 
enzyme kinetics~\cite{Agrawal2008}, but also the dynamics of {\em active} agents~\cite{Ricci:1998fj}.
Many theoretical studies have tackled these and related problems under different 
angles~\cite{Agrawal2008,Cecconi:2007dg,Dorsaz:2010zi,Fanelli:2010or,Felderhof:1976mw,Kim:2009fk,Zhou:2004oq}.
Nevertheless, a full theoretical comprehension of transport in non-ideal media remains an elusive task,
Fick's law itself and the very notion of effective diffusion coefficient being questionable  in a disordered
medium~\cite{Bouchaud:1990ys}.

In this paper we consider the subtle link between macroscopic transport equations, such as the diffusion equation,
and microscopic  processes, modeling the stochastic dynamics of some agents. The purpose of our study 
is two-fold. One the one hand, we wish to understand in greater depth the delicate procedure of obtaining 
mean-field transport equations from microscopic, agent-based stochastic models, paying special attention 
to what exactly is {\em lost} in going to the continuum. The idea is that sometimes it may prove simpler or more effective 
to describe a complex transport process (or a simple one occurring in a complex {\em milieu}) at the 
microscopic level. On the contrary, it is sometimes better to deal with macroscopic equations. 
It is thus important to investigate how the two levels of description interface with each other. 
On the other hand, as a result of our investigation, we will derive new macroscopic 
equations that provide useful analytical tools for exploring particle transport in non-ideal fluids in 
complex environments. 

The paper is organized as follows. In section~\ref{sec:1} we discuss the general framework
of simple exclusion processes (SEPs), which constitute the basic tool of our microscopic description, 
as well as the process of obtaining mean-field equations from SEPs.
In section~\ref{sec:2}, we move a step forward and consider microscopic exclusion processes involving 
agents characterized by a {\em finite} size, as opposed to standard SEPs. 
We introduce in particular the notions of {\em point-like} and {\em extended} crowding and  
discuss the formalism with reference to the problem of gravitational sedimentation. 

We then go on in section~\ref{sec:3} to propose, based
on a simple analogy, a modified diffusion equation appropriate for describing transport in non-ideal
fluids of $d$-dimensional hard spheres at high densities. The obtained equation allows one to 
recover a known result for the basic problem of particle flux at a spherical absorbing boundary in self-crowding media. 
Finally, a summary of the main results obtained in this paper is drawn in the last section.

\section{From microscopic processes to macroscopic equations\label{sec:1}}

Simple exclusion processes are space-discrete, agent-based stochastic processes modeling  
some kind of transport according to specific rules and bound to the constraint 
that no two agents can ever occupy the same site. SEPs occupy a central role in non-equilibrium 
statistical physics~\cite{Privman:1997kl,Liggett:1999fk}. While the first theoretical ideas underlying 
such processes can be traced back to Boltzmann's works~\cite{Boltzmann:1966fk}, SEPs were introduced and 
widely studied in the 70s as simplified models of one-dimensional transport for phenomena like hopping
conductivity~\cite{Richards:1977tg} and kinetics of biopolymerization [5]. Along the same lines, 
the asymmetric exclusion process (ASEP), originally introduced by Spitzer~\cite{Spitzer:1970qo}, has 
become a paradigm in non-equilibrium statistical physics~\cite{Derrida:1993bh,Schutz:1993ye,Derrida:1998oq}
and has now found many applications, such as the study of molecular motors~\cite{Golubeva:2012qf},
transport through nano-channels~\cite{Zilman:2010cr} and depolymerization of microtubules~\cite{Reese:2011nx}.

The most general SEP in one dimension is described by a stochastic jump process on a 1$D$ 
lattice with inequivalent sites in the presence of a field 
\begin{eqnarray}
\label{e:SEPMC}
\fl n_{i}(k+1) - n_{i}(k) = &z^{+}_{i-1} n_{i-1}(k)[1-n_{i}(k)] + z^{-}_{i+1} n_{i+1}(k)[1-n_{i}(k)] \nonumber \\
                            &-z^{+}_{i} n_{i}(k)[1-n_{i+1}(k)] -z^{-}_{i} n_{i}(k)[1-n_{i-1}(k)]
\end{eqnarray}
Eq.~\eref{e:SEPMC} is to be regarded as the update rule for a Monte Carlo process, where $n_{i}(k)$ is the occupancy of 
site $i$ at time $t = k \Delta t$, which can be either zero or one. 
The quantities $z^{\pm}_{i}$  are variables which have the value $0$ or $1$  according to a random number $\xi_i$ which 
has a uniform distribution between $0$ and $1$. By defining the jump probabilities $q^{\pm}_{j}$  ($j=i, i \pm 1$) one can formally write:
\begin{eqnarray}
\label{e:eta}
z^{+}_{i-1} &=& \theta(\xi_i)-\theta(\xi_i-q^{+}_{i-1}) \nonumber \\
z^{-}_{i+1} &=& \theta(\xi_i-q^{+}_{i-1})-\theta(\xi_i-q^{+}_{i-1}-q^{-}_{i+1}) \nonumber \\
z^{+}_{i} &=& \theta(\xi_i-q^{+}_{i-1}-q^{-}_{i+1})-\theta(\xi_i-q^{+}_{i-1}-q^{-}_{i+1}-q^{+}_{i}) \nonumber \\
z^{-}_{i} &=& \theta(\xi_i-q^{+}_{i-1}-q^{-}_{i+1}-q^{+}_{i})-\theta(\xi_i-1) 
\end{eqnarray}
where $\theta(\cdot)$ stands for the Heaviside step function and where we are assuming that $q^{+}_{i-1}+q^{-}_{i+1}+q^{+}_{i}+q^{-}_{i}=1$. 
Note that the ordering of appearance of the $q^{\pm}_{j}$  in the above
expressions is arbitrary. Equations~\eref{e:eta} entail that $\langle z^{\pm}_{j} \rangle = q^{\pm}_{j}$, where  $\langle \cdot \rangle$  
indicates an average over many values of $\xi_i$, for a given configuration $\{ n_{i} \}$. The above process is fully determined 
by the fields $q^{\pm}_{i}$, specifying the probability of jumping from site $i$
to site $i+1$ ($q^+_{i}$) or to site $i-1$ ($q^-_{i}$) in a time interval $\Delta t$.

A (discrete-time) master equation for the above SEP can be obtained by averaging over many Monte Carlo cycles
performed according to rule~\eref{e:SEPMC}
\begin{eqnarray}
\label{e:SEPDMEq}
\fl P_{i}(k+1) - P_{i}(k) = &q^{+}_{i-1} [P_{i-1}(k) - P_{i,i-1}(k)] + q^{-}_{i+1} [P_{i+1}(k) - P_{i,i+1}(k)] \nonumber \\
                            &-q^{+}_{i} [P_{i}(k) - P_{i,i+1}(k)] - q^{-}_{i} [P_{i}(k) - P_{i,i-1}(k)]
\end{eqnarray}
where we have defined the one-body and two-body site occupancy probabilities
\numparts\begin{eqnarray}
P_{i}(k) = \langle \langle n_{i}(k) \rangle \rangle \label{e:Poneb}\\
P_{i,i\pm 1}(k) =  \langle \langle n_{i}(k)n_{i\pm 1}(k) \rangle \rangle  \label{e:Ptwob}
\end{eqnarray}\endnumparts
Here $\langle \langle \cdot \rangle \rangle$ denotes averages performed over many independent
Monte Carlo cycles performed until time $k\Delta t$ starting from the same initial condition.
We emphasize that the same equation has been derived through 
a slightly different procedure by Richards in 1977~\cite{Richards:1977tg}.

\subsection{Mean-field equations}

With the aim of deriving macroscopic transport equations from the microscopic stochastic process described
by eqs.~\eref{e:SEPMC}, it is customary to assume a mean-field (MF) factorization,
\begin{equation}
\label{e:MFfact}
\fl P_{i,i\pm 1}(k) \equiv \langle \langle n_{i}(k)n_{i\pm 1}(k) \rangle \rangle = 
                       \langle \langle n_{i}(k) \rangle \rangle \langle \langle n_{i\pm 1}(k) \rangle \rangle = P_{i}(k) P_{i\pm 1}(k)
\end{equation}
With the help of eq.~\eref{e:MFfact}, the master equation~\eref{e:SEPDMEq} becomes
\begin{eqnarray}
\label{e:SEPMEq}
\fl P_{i}(k+1) - P_{i}(k) = &q^{+}_{i-1} P_{i-1}(k)[1 - P_{i}(k)] + q^{-}_{i+1} P_{i+1}(k)[1 - P_{i}(k)] \nonumber \\
                            &-q^{+}_{i} P_{i}(k)[1 - P_{i+1}(k)] - q^{-}_{i} P_{i}(k)[1 - P_{i-1}(k)]
\end{eqnarray}
Nonlinear mean-field equations for exclusion process of this type have been used since the 70s 
to investigate one-dimensional transport in solids~\cite{Huber:1977hc}.

Let $a$ be the lattice spacing and let us define a {\em reversal} probability $\epsilon_{i}$, such that
\begin{equation}
q^{+}_{i} = Q_{i} \qquad  q^{-}_{i} = Q_{i} - \epsilon_{i}  \label{e:field}
\end{equation}
The condition~\eref{e:field} (with $\epsilon_{i}>0$) amounts to considering a 
field introducing a bias in the positive $x$ direction.
In order to take the continuum limit $\lim_{a,\Delta t\to 0} P_{i}(k) = P(x,t)$, 
we must require 
\numparts\begin{eqnarray}
\lim_{a,\Delta t\to0} \frac{Q_{i} a^{2}}{\Delta t} = D(x) \label{e:diffc}\\
\lim_{a,\Delta t\to0} \frac{\epsilon_{i} a}{\Delta t} = v(x) \label{e:driftv}
\end{eqnarray}\endnumparts
Eq.~\eref{e:diffc} defines the position-dependent diffusion coefficient, while 
eq.~\eref{e:driftv} defines the field-induced drift velocity. Note that we are 
assuming that the reversal probability vanishes linearly with $a$.

With the help of eqs.~\eref{e:field},~\eref{e:diffc} and~\eref{e:driftv} 
it is not difficult to see that in the continuum limit eq.~\eref{e:SEPMEq} yield
\begin{equation}
\label{e:MFeq}
\frac{\partial P}{\partial t} = (1-P) \nabla^{2} [D(x) P] + D(x) P \, \nabla^{2} P - 
\frac{\partial}{\partial x} [v(x) P(1-P)] 
\end{equation}
where we have explicitly highlighted the fact that both the diffusion coefficient
and the drift velocity are position-dependent quantities.

Eq.~\eref{e:MFeq} is a nonlinear advection-diffusion equation, appropriate for describing
the continuum limit of a microscopic
exclusion process occurring on a lattice of inequivalent sites in the presence of a 
field. Although  it represents the mean-field approximation of a known master 
equation, we are not aware of any authors stating it in its most general form. 
It is interesting to note that in the case of equivalent sites, which translates to a 
constant diffusion coefficient, the diffusive part of eq.~\eref{e:MFeq} becomes linear,
{\em i.e.} the microscopic exclusion rule is {\em lost} in the diffusive part. In the case of 
zero field, one then simply recovers the ordinary diffusion equation which, as it is widely known, 
can be derived from a microscopic jump process with no exclusion rules. This curious observation has
been first reported by Huber~\cite{Huber:1977hc}. If both the diffusion coefficient and
the drift velocity are constant, eq.~\eref{e:MFeq} reduces to  
\begin{equation}
\label{e:MFeqvconst}
\frac{\partial P}{\partial t} = D\nabla^{2}P - v \frac{\partial}{\partial x} [P(1-P)] 
\end{equation}
an equation already obtained recently in Ref.~\cite{Simpson:2009ys}.

Eq.~\eref{e:MFeq} contains the single-particle probability field $P(x,t)$, which is a number between zero
and one. The value $P=1$  should correspond to the maximum density $\rho_M$ allowed in the system.
Thus, a more {\em physical} equation can be obtained by introducing the agent density
\begin{equation}
\label{e:rho}
\rho(x,t) \equiv \rho_{\scs M} P(x,t) = \frac{\phi_{\scs M}}{v_{1}(\sigma/2)} P(x,t) 
\end{equation}
where 
\begin{equation}
\label{e:v1}
v_{1}(r) = \frac{(\pi^{1/2} r)^{d}}{\Gamma(1+d/2)}
\end{equation}
is the volume of a $d$-dimensional sphere~\footnote{We emphasize that we use the general terminology of $d$-dimensional 
hard spheres. Obviously, these are hard rods in one dimension and hard disks in two.} 
of radius $r$ and $\phi_{\scs M}$ is the maximum packing fraction for systems of 
$d$-dimensional hard spheres, $\phi_{\scs M}=1$ ($d=1$), $\phi_{\scs M}=\pi/\sqrt{12}\approx 0.907$ ($d=2$) 
and $\phi_{\scs M}=\pi/\sqrt{18}\approx 0.740$ ($d=3$)~\cite{Torquato:2002kx}. 
With these definitions, and using a more general vector notation, eq.~\eref{e:MFeq} becomes
\begin{equation}
\label{e:MFeqphy}
\fl\frac{\partial \rho}{\partial t} = \left(1-\frac{\rho}{\rho_{\scs M}}\right) \nabla^{2} [D(\mathbf{x}) \rho] 
                                  + D(\mathbf{x}) \left(\frac{\rho}{\rho_{\scs M}}\right) \, \nabla^{2} \rho - 
\nabla \cdot \left[\mathbf{v}(\mathbf{x}) \rho \left(1-\frac{\rho}{\rho_{\scs M}}\right)\right] 
\end{equation}
This is the first important result of this paper.

\section{Point like versus extended crowding in one dimension\label{sec:2}}

The MF diffusion-advection equation~\eref{e:MFeq} has been derived from a master equation containing 
$exclusion$ terms of the type $(1-P_{i})$ in the limit of vanishing lattice spacing. 
This  amounts to considering agents of vanishing size in the continuum limit. We term this peculiar
situation in the macroscopic world {\em point-like crowding}. However, one may argue that 
considerable microscopic information is lost in going to the continuum limit with point-like agents.
Incidentally, this has to be the reason why the mean-field approximation does lose all the 
{\it memory} of the microscopic exclusion constraint and the
diffusion equation is recovered for equivalent
sites in the absence of a field.

Interestingly, a generalized exclusion process for agents of extended size in one dimension 
(hard rods) can be found in the literature, termed the  $\ell$-ASEP~\cite{Schonherr:2004zr}.  
The authors derive a mean-field equation, which, in the absence
of a field and for equivalent sites, reads
\begin{equation}
\label{e:LASEPdiff}
\frac{\partial \rho}{\partial t} = D\nabla^{2} \left[ 
                                              \frac{\rho}{1 - \sigma \rho}
                                            \right] 
\end{equation}
where $\sigma$ is the length of the rods. The nonlinear diffusion equation~\eref{e:LASEPdiff} 
is most interesting for a number of reasons. First of all, even if it has been derived through an ingenious but
complicate change of variables based on a quantitative mapping between the $\ell$-ASEP and the 
zero-range process~\cite{Schonherr:2004zr}, it turns out that it can be regarded as the local-density 
approximation (LDA)  of a simple general property of one-dimensional exclusion processes. 
As pointed out in 1967 by Lebowitz and Percus~\cite{Lebowitz:1967uq} concerning bulk properties 

\begin{quotation}
\noindent [$\dots$] For many purposes, however, adding a finite diameter
does not introduce any new complications; it merely
requires the replacement in certain expressions of the
actual volume per particle $\rho^{-1}$ by the reduced volume
$\rho^{-1}-\sigma$, {\em i.e.} $\rho \to \rho/(1 - \sigma \rho)$.
\end{quotation}

\noindent Remarkably, by performing the above substitution in Fick's law, under the local density 
approximation, $\rho(x,t) \to \rho(x,t)/[1 - \sigma \rho(x,t)]$, one recovers 
eq.~\eref{e:LASEPdiff}. Extending the terminology 
of the $\ell$-ASEP~\cite{Schonherr:2004zr}, we  term this scenario {\em extended-size crowding}. 
Point-like crowding in the mean field approximation corresponds to systems of fully penetrable 
spheres, while extended-size crowding yields a transport equation suitable for systems of 
totally impenetrable (hard) spheres. 

It is possible to substantiate and further illustrate the above considerations by showing that the 
transport equations for point-like and extended-size crowding for homogenous systems 
in a gravitational field allow to recover the appropriate equations of state (EOS) for 
fully penetrable spheres (FPSs) and totally impenetrable ({\em i.e.} hard) spheres (TISs), respectively.
Let us consider sedimentation in one dimension. 
Let $\varrho_{\rm p}$ and $\varrho_{\rm s}$ indicate the particle and solvent material
densities~\footnote{For the sake of clarity, we indicate with $\varrho$ mass densities and with 
$\rho$ number densities, measured in number of particles per unit volume.}, 
respectively, and $m^{\ast}= (\varrho_{\rm p}-\varrho_{\rm s})v_{\sigma}$  the buoyant mass
of the particles. The appropriate equation for point-like crowding 
bears in the case of non-zero field the signature of the microscopic exclusion process.
Recalling that $\phi_{\scs M}=1$ in one dimension and thus $\rho_{\scs M} = 1/\sigma$ 
(see eq.~\eref{e:rho}), eq.~\eref{e:MFeqphy} reduces to
\begin{equation}
\label{e:pointCsedim}
\frac{d^{2} \rho}{d z^{2}} + \frac{1}{\ell_{g}}
\frac{d}{d z} \left[\rho \left(1-\sigma\rho \right)\right] = 0
\end{equation}
Here $\ell_{g} = k_{\scs B}T/m^{\ast} g$ is the so-called sedimentation (or gravitational) length
and we have taken an upward pointing $z$-axis. Eq.~\eref{e:pointCsedim} can be solved by noting 
that at equilibrium the osmotic current is exactly balanced by the gravitational one, thus
$J = d\rho/dz + \rho(1-\sigma\rho)/\ell_{g}=0$, which gives
\begin{equation}
\label{e:sedimplc}
\rho(z)=\frac{\ds 1}{\ds \sigma\left(1 + A e^{z/\ell_{g}}\right)}
\end{equation}
The constant $A$ is to be determined by imposing the boundary conditions. Usually, this is done 
by introducing the bulk density $\rho_{0}$, such that 
\begin{equation}
\label{e:sedplcBC}
\frac{1}{h}\int_{0}^{h} \rho(z)\,dz=\rho_{0}
\end{equation}
where $h$ is the height of the initially homogenous suspension. This gives
\begin{equation}
\label{e:sedimplc1}
\rho(z)=\sigma^{-1}\left[1 +  \left( 
                                \frac{\ds 1-e^{-(1-\sigma\rho_{0})h/\ell_{g}}}
                                     {\ds e^{\sigma\rho_{0}(h/\ell_{g})}-1} 
                              \right) e^{z/\ell_{g}}\right]^{-1}
\end{equation}
For $h \gg \ell_{g}$ one has that the density at the bottom is exponentially close to the 
maximum density $\rho_{\scs M}=\sigma^{-1}$
\begin{equation}
\label{e:rhoz0plc1}
\rho(0) \simeq \frac{1}{\ds \sigma \left( 1 + e^{-\sigma\rho_{0}(h/\ell_{g})}\right)}
\end{equation}
In practice,  particle settling leaves only supernatant
solvent at the top of the cell. Hence we can safely take $h\to\infty$, which shows 
that the fluid attains maximum packing at the bottom of the cell.

The nice thing about this exercise is that we can derive an equation of state (EOS) from the 
sedimentation profile. In fact, the osmotic pressure $\Pi$  in the suspension is given by 
\begin{equation}
\label{e:EOSsedimdef}
\frac{\Pi(z)}{k_{\scs B}T} = \frac{1}{\ell_{g}}\int_{z}^{\infty} \rho(z) \, dz
\end{equation}
where, accordingly to the above considerations, we have taken the upper integration 
limit as $h\to\infty$. It is a straightforward calculation to show that the general 
expression~\eref{e:sedimplc} yields, independently of the chosen boundary conditions,
\begin{equation}
\label{e:EOSsedimplc}
\frac{\Pi}{k_{\scs B}T} = -\sigma^{-1} \log [1-\sigma\rho]
\end{equation}
The EOS~\eref{e:EOSsedimplc} has a straightforward physical interpretation. 
In a system of fully penetrable spheres with reduced density $\sigma \rho$, 
the volume fraction $\phi_{p}$ occupied by the particles 
is given by $\phi_{p}=1-\exp(-\sigma \rho)$~\cite{Torquato:2002kx}. 
Hence, the modified EOS~\eref{e:EOSsedimplc} can be obtained from the perfect gas EOS
by replacing the reduced density with its expression containing the actual volume 
fraction. This unveils the meaning of the above modified EOS, featuring the true volume fraction 
in the right-hand side. At the same time, this also illustrates the notion of point-like 
crowding.

Summarizing, a point-like microscopic exclusion process such as~\eref{e:SEPMEq}
yields a macroscopic transport equation that describes a fluid of fully penetrable spheres.  
Therefore, extending this line of reasoning
to what we have dubbed the extended-crowding scenario, we expect that the $\ell$-ASEP 
mean-field equation in non-zero field~\cite{Schonherr:2004zr} 
would yield the EOS of a system of hard rods, also known as the Tonk gas~\cite{Tonks:1936mb}
\begin{equation}
\label{e:EOSTonk}
\frac{\Pi}{k_{\scs B}T} = \frac{\rho}{1-\sigma\rho}
\end{equation}
We shall now prove that this is indeed the case. Inserting the Tonk gas EOS~\eref{e:EOSTonk} 
in eq.~\eref{e:EOSsedimdef} and differentiating with respect to $z$, one gets
\begin{equation}
\label{e:JTonk}
J_{\sigma} \equiv \frac{d \rho}{dz} + \frac{1}{\ell_{g}} \rho (1-\sigma\rho)^{2} = 0
\end{equation}
the known LDA equation for sedimentation of hard rods~\cite{Biben:1993bg}. However, 
the above equation also equals the condition that the total $\ell$-ASEP
particle flux $J_{\sigma}$ vanishes at equilibrium. Indeed, the 
stationary mean-field equation corresponding to the $\ell$-ASEP 
reads~\cite{Schonherr:2004zr} 
\begin{equation}
\label{e:LASEPdifffield}
\frac{d^{2}}{dz^{2}}\left[  
               \frac{\rho}{1 - \sigma \rho}
            \right] + \frac{1}{\ell_{g}} \frac{d \rho}{dz} = 0 
\end{equation}
which leads to the same expression for the total (constant) particle flux as eq.~\eref{e:JTonk}.
Hence, the $\ell$-ASEP  yields a macroscopic transport equation that describes 
a gas of hard rods in the local density approximation. 

The discussion above which leads to the concept of {\em extended} crowding  
applies to one dimensional systems. Starting from this setting, one can 
raise the question whether similar arguments might be 
employed to obtain a modified nonlinear equation accounting for excluded volume effects in the diffusion 
of hard spheres in two and three dimensions. 
The following section is devoted to speculating further along this line of reasoning.

%
\section{Excluded volume interactions of finite-size agents in higher dimensions.}
\label{sec:3}

In the preceding section we have shown that  the {\em point-like} crowding scenario
leads to the mean-field equation~\eref{e:MFeqphy} from a standard simple exclusion process 
at the {\em microscopic} level.  
The {\em extended-crowding} generalization consists in endowing agents with a finite size (hard 
core). The ensuing exclusion process in one dimension has been
termed in the literature the $\ell$-ASEP~\cite{Schonherr:2004zr}.
%
\begin{table}[t!]
\centering
\caption{
\label{t:tabCPDF} The particle conditional pair distribution function $G_{p}$ for stationary 
ensembles of $d$-dimensional spheres of diameter $\sigma$~\cite{Torquato:2002kx}}
\begin{tabular}{@{}l c c c}
\br & $d=1$ & $d=2$ & $d=3$ \\
\mr
Fully penetrable spheres & 1 & 1 & 1 \\
Totally impenetrable spheres & $\frac{\ds 1}{\ds 1-\sigma \rho}$ & $a_{0} + a_{1}\left(\frac{\ds \sigma}{\ds r}\right)$ &
$b_{0} + b_{1}\left(\frac{\ds \sigma}{\ds r}\right) + b_{2}\left(\frac{\ds \sigma}{\ds r}\right)^{2}$ \\
\br
\end{tabular}
\end{table}

We have shown that the nonlinear transformation put forward by Lebowitz and Percus in 1967, 
under the local density approximation, allows one to recover the $\ell$-ASEP mean-field 
transport equation~\eref{e:LASEPdiff} for the evolution of the density of extended 
rod-like agents in one dimension. 
Interestingly, it is possible to posit a generalization of such macroscopic 
transport equation adequate to the mean-field limit of exclusion processes involving 
extended objects in more than one dimension
by borrowing general concepts in the theory of heterogenous media.

With reference to standard definitions of micro-structural descriptors in 
$d$ dimensions~\cite{Torquato:2002kx}, we can identify the aforementioned substitution by Lebowitz and Percus 
as a mapping between certain statistical properties characterizing systems of fully penetrable spheres and 
totally impenetrable ({\em i.e.} hard) spheres. 
More precisely, let us introduce the so-called
{\em conditional pair distribution function} (CPDF)  $G_p(r)$. Let $r$ denote the distance from the 
center of some reference particle in a system with bulk density $\rho_{0}$. 
Then, by definition $\rho_{0} s_{1}(r)G_{p}(r) \, dr$ 
equals the average number of particles in the shell of infinitesimal volume $s_{1}(r)\, dr$ around 
the central particle, given that the volume $v_{1}(r)$ of the $d$-sphere of radius $r$ is empty 
of other particle centers. Here
\begin{equation}
\label{e:v1s1}
s_{1}(r) = \frac{dv_{1}(r)}{dr} = \frac{2\pi^{d/2}r^{d-1}}{\Gamma(d/2)}
\end{equation}
The CPDF for FPS and TIS systems are reported in Table~\eref{t:tabCPDF},
where one can readily recognize the Lebowitz and Percus substitution as a mapping 
between the FPS and TIS CPDFs in one dimension.
The coefficients appearing in the expressions of $G_{p}(r)$ are given by the 
following expressions~\footnote{These are the expressions appropriate to 
the liquid phase below the freezing point, 
{\em i.e.} for $\phi < \phi_{f}$ ($\phi_{f}=0.69$ for $d=2$ and 
$\phi_{f}=0.494$ for $d=3$).} for $d=2$~\cite{Torquato:2002kx}
\numparts\begin{eqnarray}
a_{0} = \frac{1 + 0.128\, \phi}{(1-\phi)^{2}}  \label{e:a0}\\
a_{1} = -\frac{0.564\,\phi}{(1-\phi)^{2}}      \label{e:a1}
\end{eqnarray}\endnumparts
while for $d=3$ one has
\numparts\begin{eqnarray}
 b_{0} = \frac{1+\phi+\phi^{2}-\phi^{3}}{(1-\phi)^{3}}     \label{e:b0}\\
 b_{1} = \frac{\phi(3\phi^{2}-4\phi-3)}{2(1-\phi)^{3}}     \label{e:b1}\\
 b_{2} = \frac{\phi^{2}(2-\phi)}{2(1-\phi)^{3}}            \label{e:b2}
\end{eqnarray}\endnumparts
where $\phi = v_{1}(\sigma/2)\rho$ is the packing fraction,  $\sigma$ being 
the diameter of a $d$-dimensional sphere. 

The above discussion suggests a way to generalize the $\ell$-ASEP~\eref{e:LASEPdiff} to describe 
excluded-volume effects in more than one dimension in the mean-field approximation in a homogeneous medium
and zero field. For a spherically symmetric problem we posit 
\begin{equation}
\label{e:LASEP23dr}
\frac{\partial \rho}{\partial t} = D\nabla^{2} [\rho \, G_{p}( \rho,r)]
\end{equation}
where we have emphasized that $G_{p}$ depends explicitly on $r$ (for $d>1$). 
Eq.~\eref{e:LASEP23d} is the second main result of this paper.
The coefficients appearing in the expressions for $G_{p}$ are given by 
eqs~\eref{e:a0},~\eref{e:a1},~\eref{e:b0},~\eref{e:b1} 
and~\eref{e:b2} and it is understood that, according to the local density approximation, 
one should replace $\phi$ with $v_{\sigma} \rho(r,t)$, 
where $v_{\sigma}\equiv v_{1}(\sigma/2)$ is the volume of one hard $d$-dimensional sphere,
so that $G_{p}$ depends on $r$ both explicitly and implicitly through $\rho(r)$.

\subsection{The Smoluchowski problem at finite densities}
In order to provide some {\em a posteriori} justification for  eq.~\eref{e:LASEP23dr},
it is instructive to consider how  the classical problem of diffusion to an absorbing sphere is modified
in a non-ideal fluid. Let us imagine a fixed sink 
of radius $R_{s}$ that absorbs hard spheres of radius $\sigma/2$ and bulk density $\rho_{0}$.
The rate $k$ measuring the number of particles absorbed by the sink per unit time  
equals the total flux into the sink. For ordinary Fickian diffusion, one has the classical 
result $k = k_{\scs S} \equiv 4 \pi D (R_{s}+\sigma/2) \rho_{0}$, known as the Smoluchowski rate~\cite{Smoluchowski:1916fk}.
This result is indeed the prediction of a two-body problem, {\em i.e.} it amounts to considering the absorption of
non-interacting, or equivalently fully penetrable, spheres. Thus, it describes the problem in the 
infinite dilution limit. Eq.~\eref{e:LASEP23dr} can now be employed to repeat the same exercise for hard 
spheres at finite densities,
that is, the Smoluchowski problem with excluded-volume interactions accounted for. One should then solve 
the following boundary-value problem 
\numparts\begin{eqnarray}
\nabla^{2} [\rho\, G_{p}(v_{\sigma}\rho,r)] = 0 \label{e:BVPGp}\\
\rho(r=R_{s} + \sigma/2) = 0 \label{e:BC1}\\
\lim_{r\to\infty} \rho(r) = \rho_{0} \label{e:BC2}
\end{eqnarray}\endnumparts
The rate can be computed readily without really solving the (modified) Laplace equation. From~\eref{e:BVPGp}, we
have directly
\begin{equation}
\label{e:rate1}
\frac{\partial}{\partial r} \left[ \rho(r) G_{p}(\phi(r),r)\right] = \frac{k}{4 \pi D r^{2}}
\end{equation}
where we have defined $\phi(r)=v_{\sigma}\rho(r)$, so that $\phi=v_{\sigma}\rho_{0}$ denotes the bulk packing fraction
of the hard spheres. Integrating eq.~\eref{e:rate1} between $R_{s}+\sigma/2$ and infinity 
and taking into account the boundary conditions~\eref{e:BC1}
and~\eref{e:BC2}, it is straightforward to obtain 
\begin{equation}
\label{e:rate2}
\frac{k}{k_{\scs S}} = G_{p}^{\infty}
\end{equation}
where $G_{p}(\infty)\equiv\lim_{r\to\infty}G_{p}(\phi(r),r)$. In three dimensions, one thus 
has $k/k_{\scs S}=b_{0}(\phi)$, where one can recognize $b_{0}(\phi)$ 
as the compressibility $Z(\phi)$ of the hard sphere fluid in the Carnahan-Starling 
approximation~\cite{Carnahan:1969ys}. 
We see that eq.~\eref{e:LASEP23dr} allows one 
to recover our previous result $k/k_{\scs S}=Z(\phi)$, obtained in two different ways,  
by assuming a density-dependent mobility in the diffusion equation~\cite{Dorsaz:2010vn} and
from a transport equation derived in the local-density approximation~\cite{Piazza:2013fk}.

\section{Summary and discussion}
%

In this paper we have discussed a general framework allowing to 
obtain macroscopic transport equation accounting 
for excluded volume effects in systems of $d$-dimensional hard spheres starting 
from a microscopic stochastic exclusion processes. The aim of this procedure is to derive mean-field equations
suitable for describing transport processes in many-body systems in highly non-ideal conditions. 
We have identified two strategies for doing so. The first route, termed {\em point-like} crowding,
leads from a standard simple exclusion process to the mean-field equation~\eref{e:MFeqphy}. 
For a homogenous medium in the absence of a field this reduces to a simple diffusion equation,
which is why we have dubbed this scenario {\em point-like} crowding. Only for inequivalent sites
and/or in the presence of a field does the microscopic exclusion constraint survives in the mean-field limit.
The second strategy, named {\em extended-crowding}, takes inspiration from a modified microscopic 
exclusion process in one dimension involving extended agents, the so-called $\ell$-ASEP~\cite{Schonherr:2004zr}.
By extending an argument originally put forward by Lebowitz and Percus in 1967, coupled to a local
density approximation, we have posited a modified nonlinear diffusion equation suitable for 
studying in an effective manner transport processes in dense systems of hard spheres, eq.~\eref{e:LASEP23d}.
We have shown that this equation allows one to recover recent results obtained for the 
problem of diffusion to an absorbing sphere in a self-crowded medium.    
Furthermore, we have brought to the fore an interesting structure underlying the two above strategies for
obtaining macroscopic equations from microscopic stochastic processes. While extended crowding 
is indeed appropriate for diffusion of hard spheres, the continuum limit of point-like crowding
exclusion processes is only appropriate for systems of fully penetrable spheres. 
Therefore, when one takes the continuum limit, all signatures of the microscopic exclusion constraints 
are lost in the diffusive part of the transport equation.

We note that eq.~\eref{e:LASEP23dr} provides a sensible description 
of diffusion in a non-ideal fluid in the case where the problem has spherical symmetry.
We may ask whether our line of reasoning may be extended to a problem with 
different or no symmetry. Guided by the observation that $G_{p}^\infty = Z(\phi)$ 
for a system of $d$-dimensional hard spheres, we can speculate that a modified
equation could be considered, involving the {\em bulk}  
CPDF $G_{p}^{\infty}(\phi)$  in the local density approximation,
\begin{equation}
\label{e:LASEP23d}
\frac{\partial \rho}{\partial t} = D\nabla^{2} [\rho \, G^{\infty}_{p}( \rho)]
\end{equation}
where again one should understand  $\phi \to v_{\sigma} \rho(r,t)$ in the expression for 
the  coefficients $a_{i}(\phi)$ and $b_{i}(\phi)$. Some justification for the above idea can be 
gathered by recalling the known equation that one obtains by 
introducing a density-dependent mobility in the diffusion equation
given by the derivative of the osmotic pressure with respect to the density~\cite{Dhont1996},
$D(\rho) = D_{0} \beta \, d\Pi(\rho)/d \rho = D_{0} Z(\rho)$, namely
\begin{equation}
\label{e:dhont}
\frac{\partial \rho}{\partial t} = D_{0} \nabla \cdot [ Z(\rho) \nabla \rho)]
\end{equation}
where $D_{0}$ is the {\em bare} diffusion coefficient (corresponding to the infinite dilution 
limit) and $Z(\rho)$ is the compressibility factor. At zero order, where $G_{p}(\rho) = G_{p}(\rho_{0}) = const.$
and $Z(\rho) = Z(\rho_{0}) = const.$ eq.~\eref{e:LASEP23d} and eq.~\eref{e:dhont} are the same,
as $G_{p}^{\infty}(\phi) = Z(\phi)$. 

As a last observation, we see that, if we compare eq.~\eref{e:LASEPdifffield} with eq.~\eref{e:pointCsedim},  
the point-like route to a macroscopic mean-field equation for a homogenous medium bears the signature of the microscopic
exclusion mechanism in the advection term. On the contrary, at least in one dimension, 
the extended-crowding procedure yields a modified  diffusion term, while the term associated with 
the external field current is left unchanged. The profound meaning of this fact appears still elusive, but it 
would be certainly interesting to dig further.

%
\bigskip
\indent The authors would like to thank L. E. L. Kann and G. Foffi for a critical reading of 
the manuscript and enlightening discussions.
%
%

\begin{thebibliography}{10}
\expandafter\ifx\csname url\endcsname\relax
  \def\url#1{{\tt #1}}\fi
\expandafter\ifx\csname urlprefix\endcsname\relax\def\urlprefix{URL }\fi
\providecommand{\eprint}[2][]{\url{#2}}

\bibitem{Crank:uq}
Crank J 1975 {\em The mathematics of diffusion\/} (New York: Oxford University
  Press)

\bibitem{Dhont1996}
Dhont J 1996 {\em An introduction to dynamics of colloids\/} (Amsterdam:
  Elsevier)

\bibitem{McGuffee:2010tg}
McGuffee S~R and Elcock A~H 2010 {\em PLoS Comput Biol\/} {\bf 6} e1000694 EP
  --

\bibitem{Caspi:2002zr}
Caspi A, Granek R and Elbaum M 2002 {\em Phys. Rev. E\/} {\bf 66}(1) 011916

\bibitem{kremp:216}
Kremp D, Schlanges M, Bonitz M and Bornath T 1993 {\em Physics of Fluids B:
  Plasma Physics\/} {\bf 5} 216--229

\bibitem{Saffman01081975}
Saffman P~G and Delbr\"uck M 1975 {\em Proceedings of the National Academy of
  Sciences\/} {\bf 72} 3111--3113

\bibitem{Talmon:2010bu}
Talmon Y, Shtirberg L, Harneit W, Rogozhnikova O~Y, Tormyshev V and Blank A
  2010 {\em Phys. Chem. Chem. Phys.\/} {\bf 12}(23) 5998--6007

\bibitem{Mitra:1992pt}
Mitra P~P, Sen P~N, Schwartz L~M and Le~Doussal P 1992 {\em Phys. Rev. Lett.\/}
  {\bf 68}(24) 3555--3558

\bibitem{Whitaker:1967ta}
Whitaker S 1967 {\em AIChE Journal\/} {\bf 13} 420--427

\bibitem{Dix:2008lp}
Dix J~A and Verkman A 2008 {\em Annual Review of Biophysics\/} {\bf 37}
  247--263

\bibitem{Phair:2000wa}
Phair R~D and Misteli T 2000 {\em Nature\/} {\bf 404} 604--609

\bibitem{Condamin:2007ez}
Condamin S, Benichou O, Tejedor V, Voituriez R and Klafter J 2007 {\em
  Nature\/} {\bf 450} 77--80

\bibitem{Zhou:2004km}
Zhou H~X 2004 {\em Journal of Molecular Recognition\/} {\bf 17} 368--375

\bibitem{Minton2000a}
Minton A~P 2000 {\em Curr Opin Struct Biol\/} {\bf 10} 34--39

\bibitem{Zimmerman1993}
Zimmerman S~B and Minton A~P 1993 {\em Annu Rev Biophys Biomol Struct\/} {\bf
  22} 27--65

\bibitem{Kim:2009fk}
Kim J~S and Yethiraj A 2009 {\em Biophysical journal\/} {\bf 96} 1333--1340

\bibitem{Ando:2010kl}
Ando T and Skolnick J 2010 {\em Proceedings of the National Academy of
  Sciences\/} {\bf 107} 18457--18462

\bibitem{Balbo:2013ov}
Balbo J, Mereghetti P, Herten D~P and Wade R~C 2013 {\em Biophysical Journal\/}
  {\bf 104} 1576 -- 1584

\bibitem{Weiss:2004vn}
Weiss M, Elsner M, Kartberg F and Nilsson T 2004 {\em Biophysical journal\/}
  {\bf 87} 3518--3524

\bibitem{ben-Avraham:2000kx}
ben Avraham D and Havlin S 2000 {\em Diffusion and Reactions in Fractals and
  Disordered Systems\/} (Cambridge University Press)

\bibitem{Bouchaud:1990ys}
Bouchaud J~P and Georges A 1990 {\em Physics Reports\/} {\bf 195} 127--293

\bibitem{Novak2009758}
Novak I~L, Kraikivski P and Slepchenko B~M 2009 {\em Biophysical Journal\/}
  {\bf 97} 758 -- 767

\bibitem{Konopka01092006}
Konopka M~C, Shkel I~A, Cayley S, Record M~T and Weisshaar J~C 2006 {\em
  Journal of Bacteriology\/} {\bf 188} 6115--6123

\bibitem{Nicholson01121981}
Nicholson C and Phillips J~M 1981 {\em The Journal of Physiology\/} {\bf 321}
  225--257

\bibitem{Smoluchowski:1916fk}
von Smoluchowski M 1916 {\em Physik Z\/} {\bf 17} 557--571

\bibitem{Zhou:2004oq}
Zhou H~X 2004 {\em Journal of Molecular Recognition\/} {\bf 17} 368--375

\bibitem{Cheung2005}
Cheung M~S, Klimov D and Thirumalai D 2005 {\em Proc Natl Acad Sci U S A\/}
  {\bf 102} 4753--4758

\bibitem{Dzubiella2005}
Dzubiella J and McCammon J~A 2005 {\em The Journal of Chemical Physics\/} {\bf
  122} 184902 (pages~7)

\bibitem{Piazza2005}
Piazza F, De~Los~Rios P, Fanelli D, Bongini L and Skoglund U 2005 {\em European
  Biophysics Journal\/} {\bf 34} 899--911

\bibitem{Fanelli:2010or}
Fanelli D and McKane A~J 2010 {\em Physical Review E\/} {\bf 82} 021113

\bibitem{Schmit2009}
Schmit J~D, Kamber E and Kondev J 2009 {\em Physical Review Letters\/} {\bf
  102} 218302

\bibitem{Tachiya2007}
Tachiya M and Traytak S~D 2007 {\em Journal of Physics: Condensed Matter\/}
  {\bf 19} 065111

\bibitem{Agrawal2008}
Agrawal M, Santra S~B, Anand R and Swaminathan R 2008 {\em PRAMANA - Journal of
  Physics\/} {\bf 71} 359--368

\bibitem{Ricci:1998fj}
Ricci N, Barbanera F and Erra F 1998 {\em The Journal of Eukaryotic
  Microbiology\/} {\bf 45} 606--611

\bibitem{Cecconi:2007dg}
Cecconi F, De~Los~Rios P and Piazza F 2007 {\em Journal of Physical Chemistry
  B\/} {\bf 111} 11057--11063

\bibitem{Dorsaz:2010zi}
Dorsaz N, De~Michele C, Piazza F and Foffi G 2010 {\em Journal of Physics:
  Condensed Matter\/} {\bf 22}

\bibitem{Felderhof:1976mw}
Felderhof B~U and Deutch J~M 1976 {\em Journal of Chemical Physics\/} {\bf 64}
  4551--4558

\bibitem{Privman:1997kl}
Privman V (ed) 1997 {\em Nonequilibrium Statistical Mechanics in One
  Dimension\/} (Cambridge University Press.)

\bibitem{Liggett:1999fk}
Liggett T~M 1999 {\em {Stochastic Interacting Systems: Contact, Voter and
  Exclusion Processes}\/} (Berlin: Springer-Verlag)

\bibitem{Boltzmann:1966fk}
Boltzmann L 1896 {\em Vorlesungen \"uber Gastheorie (Lectures on Gas Theory)\/}
  vol I and II (Berkeley: University of California Press. Translated by S. G.
  Brush, 1966)

\bibitem{Richards:1977tg}
Richards P~M 1977 {\em Physical Review B\/} {\bf 16} 1393--1409

\bibitem{Spitzer:1970qo}
Spitzer F 1970 {\em Advances in Mathematics\/} {\bf 5} 246--290

\bibitem{Derrida:1993bh}
Derrida B, Evans M~R, Hakim V and Pasquier V 1993 {\em Journal of Physics A:
  Mathematical and General\/} {\bf 26} 1493

\bibitem{Schutz:1993ye}
Sch{\"u}tz G and Domany E 1993 {\em Journal of Statistical Physics\/} {\bf 72}
  277--296

\bibitem{Derrida:1998oq}
Derrida B 1998 {\em Physics Reports\/} {\bf 301} 65--83

\bibitem{Golubeva:2012qf}
Golubeva N and Imparato A 2012 {\em Physical Review Letters\/} {\bf 109} 190602

\bibitem{Zilman:2010cr}
Zilman A and Bel G 2010 {\em Journal of Physics: Condensed Matter\/} {\bf 22}
  454130

\bibitem{Reese:2011nx}
Reese L, Melbinger A and Frey E 2011 {\em Biophysical journal\/} {\bf 101}
  2190--2200

\bibitem{Huber:1977hc}
Huber D~L 1977 {\em Physical Review B\/} {\bf 15} 533--538

\bibitem{Simpson:2009ys}
Simpson M~J, Landman K~A and Hughes B~D 2009 {\em Physical Review E\/} {\bf 79}
  031920--

\bibitem{Torquato:2002kx}
Torquato S 2002 {\em Random Heterogenous Materials\/} Interdisciplinary Applied
  Mathematics (New York: Springer-Verlag)

\bibitem{Schonherr:2004zr}
Sch{\"o}nherr G and Sch{\"u}tz G~M 2004 {\em Journal of Physics A: Mathematical
  and General\/} {\bf 37} 8215

\bibitem{Lebowitz:1967uq}
Lebowitz J~L and Percus J~K 1967 {\em Physical Review\/} {\bf 155} 122--138

\bibitem{Tonks:1936mb}
Tonks L 1936 {\em Physical Review\/} {\bf 50} 955--963

\bibitem{Biben:1993bg}
Biben T, Hansen J~P and Barrat J~L 1993 {\em The Journal of Chemical Physics\/}
  {\bf 98} 7330--7344

\bibitem{Carnahan:1969ys}
Carnahan N and Starling K 1969 {\em J.~Chem.~Phys.\/} {\bf 51} 635--636

\bibitem{Dorsaz:2010vn}
Dorsaz N, De~Michele C, Piazza F, De~Los~Rios P and Foffi G 2010 {\em Physical
  Review Letters\/} {\bf 105} 120601--

\bibitem{Piazza:2013fk}
Piazza F, Foffi G, De~Michele C and De~Los~Rios P 2013 {\em submitted to:
  Journal of Physics: Condensed Matter\/}

\end{thebibliography}
%
\providecommand{\newblock}{}

%
%
\end{document}